%% file: paper.tex
\documentclass[sigconf, nonacm]{acmart}

\usepackage{adjustbox}
\usepackage{xspace}
\usepackage[noabbrev,capitalize,nameinlink]{cleveref}

\usepackage[inline]{aplcomments}

\begin{document}
\title{Cost-Intelligent Data Analytics in the Cloud}

\author{Huanchen Zhang}
\affiliation{%
  \institution{Tsinghua University}
}
\email{huanchen@tsinghua.edu.cn}

\author{Yihao Liu}
\affiliation{%
  \institution{Tsinghua University}
}
\email{liuyihao21@mails.tsinghua.edu.cn}

\author{Jiaqi Yan}
\affiliation{%
  \institution{Snowflake}
}
\email{jiaqi@snowflake.com}

\include{abstract}
\maketitle

\input{intro}

\input{motivation}

\input{resource}

\input{tuning}

\input{concl}

\bibliographystyle{ACM-Reference-Format}
\bibliography{paper}

\end{document}

%% file: abstract.tex
\begin{abstract}
\label{sec:abstract}

For decades, database research has focused on optimizing performance
under fixed resources.
As more and more database applications move to the public cloud,
we argue that it is time to make cost a first-class citizen when
solving database optimization problems.
In this paper, we introduce the concept of cost intelligence
and envision the architecture of a cloud data warehouse designed for that.
We investigate two critical challenges to achieving cost intelligence
in an analytical system:
automatic resource deployment and cost-oriented auto-tuning.
We describe our system architecture with an emphasis on the components
that are missing in today's cloud data warehouses.
Each of these new components represents unique research opportunities
in this much-needed research area.

\end{abstract}

%% file: intro.tex
\section{Introduction}
\label{sec:intro}

From an economic perspective, databases are simply a type of goods or services.
A consumer pays a price of $C$ for the database product and earns utility (which
eventually translates to revenue) by receiving timely query results.
The utility function $U(p)$ is positively correlated with the query performance $p$.
A ``better'' (i.e., more competitive in the market) database, therefore,
is the one that maximizes consumer profit (i.e., $\Pi = U(p) - C$).
This fundamental logic has always remained the same throughout database development.

For decades, database research has focused on optimizing the performance $p$ under a fixed
amount of resources, leaving the cost $C$ behind.
This is rational because traditional distributed databases~\cite{lamb2012vertica,lyu2021greenplum,teradata}
typically dedicate a predetermined number of machines to run the service. 
The dominating factor in $C$ is the sunk cost $C_{sunk}$ for purchasing computing equipment,
software licenses, and other supporting facilities.
The marginal cost $\Delta C$ attributed to database operations (e.g., power consumption)
is relatively small.
There is little one can do to increase the user profit by reducing the cost.

The situation has changed dramatically in the past decade as more and more database applications
move to the public cloud.
The ``pay-as-you-go'' pricing model eliminates the majority of the sunk cost and enables
fine control over the operational cost $\Delta C$.
Technologies such as the disaggregation of compute and storage allow recent cloud-native
databases~\cite{dageville2016snowflake,armenatzoglou2022reshiftnew,bigquery}
to further leverage the resource elasticity at a fine granularity.
Meanwhile, cost reduction is the driving force behind most customers' migration to
cloud-native database services.
Performance is no longer the only criterion.
A typical database user today treats performance as a requirement rather than an optimization
target~\cite{ortiz2015sla,florescu2009} 
because the performance beyond often contributes little to the application's revenue
(i.e., $U(p)$ is a step function).
Their goal is to minimize the cost while guaranteeing a performance service-level agreement (SLA).
Another user paradigm is to set a fixed budget to spend on the database service and try to
get the best performance out of it.

We argue that it is time to make \textbf{cost} a first-class citizen when solving database
optimization problems.
The demand for cost savings has always been there, and the resource elasticity provided
by the public cloud services
makes it possible to manipulate cost as a free variable rather than a near-constant.
Cost efficiency is equally important as performance because they both serve to maximize user profit.
Given this, optimization in a cloud-native database should be bi-objective by default:
any performance gain must be justified by the potential trade-off in cost (and vice versa)
to be considered valuable in a specific application.

It is important to distinguish between user-observable cost (UOC) and
provider-observable cost (POC, often referred to as the cost of goods sold financially).
From a user's perspective, their cost is the cloud bill which reflects the amount of
resources \textit{reserved/promised} by the database service for completing a task.
A simple aggregation of the UOCs, however, does not equal the cost
borne by the service provider (i.e., POC).
To support multi-tenancy, a provider typically manages virtualized resource pools
where smart scheduling algorithms (sometimes overcommitting) could lead to a much higher
resource utilization overall compared to that of individual users~\cite{narasayya2021survey}.
To this end, UOC is the base cost that determines how cost-competitive a database product is, 
while optimizing POC further improves the service provider's profit margin.
The focus of this paper is on optimizing UOC, that is,
reducing the resource requirement and waste for completing a user query while
guaranteeing its performance SLA.
This is primarily orthogonal to the multi-tenancy techniques for POC optimization,
and providers are willing to spend effort reducing UOC to stay cost-competitive in the market.
Techniques for improving the efficiency of multi-tenancy in a data center~\cite{narasayya2021survey}
is beyond the scope of this paper.

The idea of treating monetary cost as a database optimization target
dates back to the early days of cloud computing~\cite{florescu2009}.
There is a rich literature on optimizing the resource allocation
in big data systems with massive parallelism~\cite{pimpley2022towards,yint2018,siddiqui2020cost,ferguson2012jockey,kllapi2011schedule,lyu2022fine}.
Recent studies on cost-efficient cloud data warehouses concentrate mainly
on cloud configurations.
For example, Leis and Kuschewski proposed a model-based algorithm
to select a cost-optimal instance configuration to run a workload~\cite{leis2021towards}.
Tan et al. examined major cloud OLAP engines and revealed the performance-cost
trade-offs in the cloud storage hierarchy (e.g., AWS S3 vs. EBS)~\cite{tan2019}.
Starling~\cite{perron2020starling} and Lambada~\cite{muller2020lambada}
used cloud functions to execute queries to save cost by avoiding resource over-provisioning.
These solutions, however, are ad-hoc, and they target only a single aspect
of achieving cost efficiency in a cloud analytics system.
There is a need for a holistic design of the core database architecture
that treats performance and cost with equal importance
and allows a native bi-objective optimization.
The reality is that cost control is still difficult in state-of-the-art systems,
and the burden of cost management is mostly left to the users~\cite{rajan2016perforator, pimpley2022towards}. 

In this paper, we envision the architecture of a cloud data warehouse
designed for cost intelligence.
We begin by discussing the cost-control challenges in today's cloud data warehouses.
We then present the concept of cost intelligence
and describe our architectural design for solving the critical problems in
automatic resource deployment in the foreground
and automatic database tuning in the background.
The paper serves as both a system blueprint and a research roadmap
that identifies the missing/suboptimal components and algorithms
for achieving cost intelligence in the next-generation cloud data warehouse.

%% file: motivation.tex
\section{The Case For Cost Intelligence}
\label{sec:motivation}
\begin{figure}[t!]
\begin{minipage}{0.32\columnwidth}
\centering
\includegraphics[width=\columnwidth]{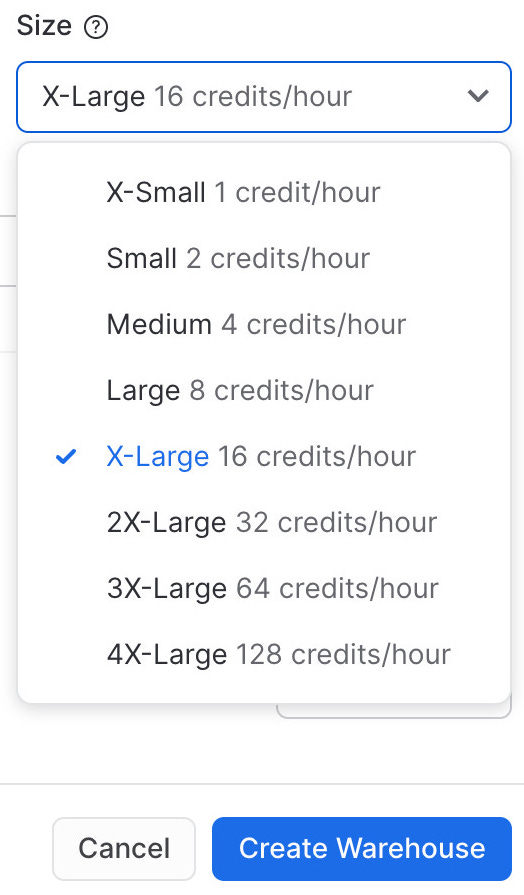}
\caption{Snowflake User Interface}
\label{fig:ui}
\end{minipage}
\begin{minipage}{0.66\columnwidth}
\centering
\includegraphics[width=\columnwidth]{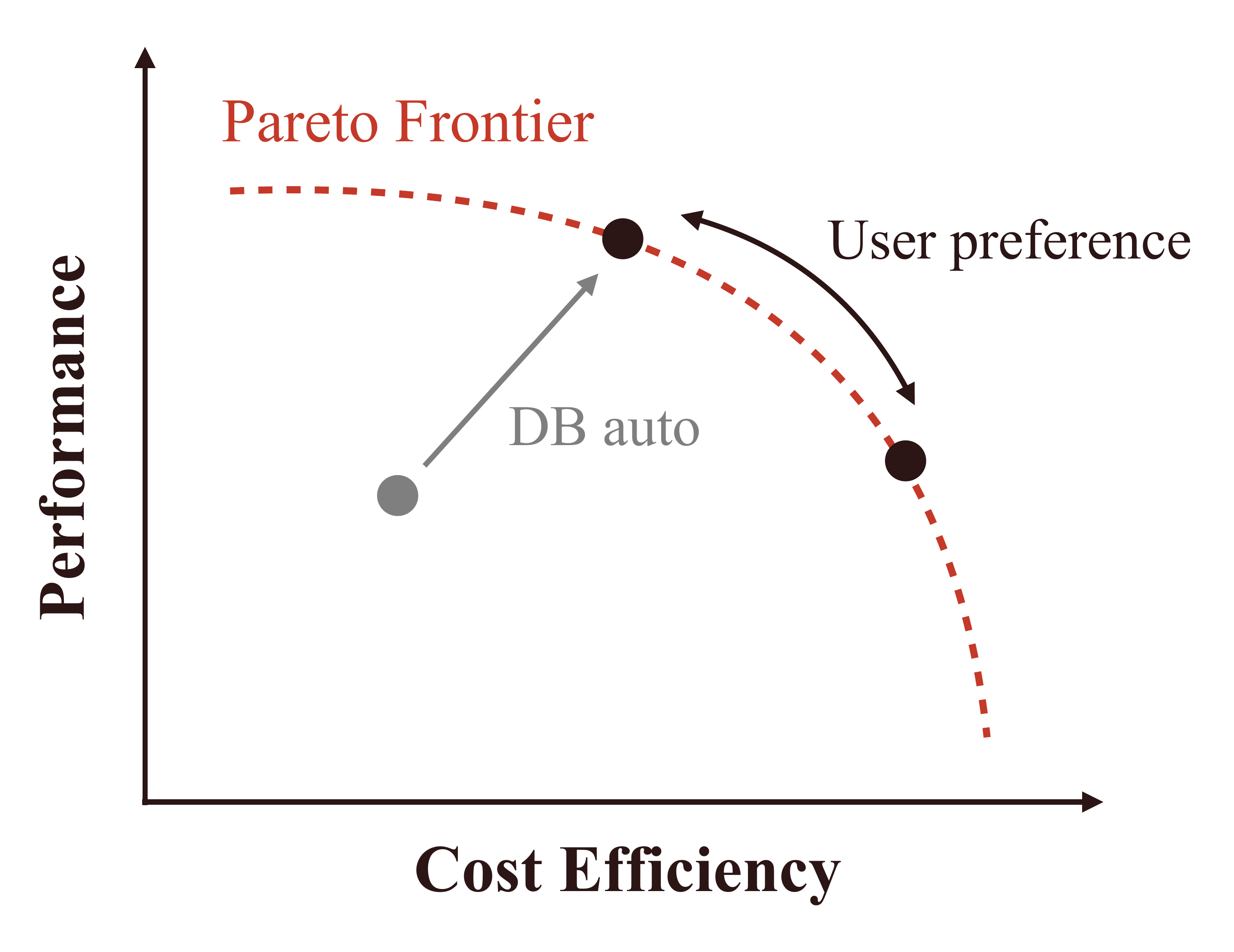}
\caption{Cost Intelligence}
\label{fig:cost-intel}
\end{minipage}
\end{figure}

Despite the ``pay-as-you-go'' model in the public cloud,
it is still difficult for an average database user to leverage resource elasticity
to control and optimize their bills when using a cloud database service.
In this section, we identify the cost-optimization challenges in existing systems
for both online resource provisioning and offline database tuning.
These challenges make the case for a cost-intelligent database design that frees
the users from the burden of pursuing cost efficiency manually and ad hoc.

The first key challenge is \textbf{automatic resource deployment} during query execution.
\cref{fig:ui} shows a partial user interface (UI) for creating a virtual warehouse
(i.e., a stateless cluster for query processing) in Snowflake~\cite{dageville2016snowflake}.
Before submitting any queries, a user must determine the cluster size by choosing a predefined ``T-shirt'' size, where a larger size means more computing nodes and a more expensive unit price.
This basic service model for resource allocation is common in today's cloud data warehouses.
However, this ``one-shot'' user-provisioning model often leads to inefficient resource utilization.

First, average database users lack the expertise to accurately estimate the resource necessary for their workloads.
As a result, they often \emph{over-provision} the cluster size to guarantee that their performance SLAs are met.
Second, the cluster size is predetermined and \emph{fixed}\footnote{although some services
support auto-scaling to handle workload bursts~\cite{armenatzoglou2022reshiftnew}.} for the entire workload,
regardless of the query complexity and data volume changes during the execution.
Such inflexibility prevents each pipeline within an analytical query from reaching its cost-optimal degree of parallelism (DOP).

Determining the cost-optimal DOP for each pipeline in a distributed query plan is
a pivotal step toward automatic resource deployment.
An interesting logic enabled by the resource elasticity in the cloud is that
for a task that is embarrassingly parallel, executing the task using 1 machine
for 100 minutes incurs the same dollar cost as executing the task using 100 machines
for 1 minute, but the second configuration has a 100x performance advantage.
However, allocating more machines does not always bring performance boosts for free
because most database operators do not exhibit perfectly-linear scalability.
Many of them (e.g., hash partitioning) require exchanging data between the machines
where the network could become the system's bottleneck.
Unlike in a map-reduce-based big data system~\cite{pimpley2022towards,venkataraman2016ernest},
over-scaling the cluster size in a distributed database not only wastes resources
but also could have a negative impact on query latency.
A user may end up paying more for the same or even worse query performance.

The second cost-optimization challenge is \textbf{automatic database tuning} in the background.
The goal is to allow databases to apply tuning actions (e.g., building indexes and materialized views)
wisely and automatically to cut the expenses on database administrators (DBAs).
After decades of research and practice, from the AutoAdmin project~\cite{chaudhuri1998autoadmin} to the recent
self-driving databases~\cite{pavlo2017self},
there are sophisticated algorithms and tools that can
propose tuning actions beneficial to the overall system performance.

These auto-tuning tools, however, are not designed to be cost-aware in the cloud environment,
and they do not provide a customer-understandable measure that can clearly indicate the net gain
(or loss) of a particular tuning action.
For example, suppose that a user is presented with a tuning suggestion that proposes to
recluster (or repartition) a petabyte-sized table \texttt{T} according to a different attribute \texttt{A}.
Although such a reclustering operation could speed up queries that use \texttt{A} in the
predicates or join columns, the cost of repopulating a petabyte-sized table is enormous.
Without a metric to evaluate the pros and cons of a tuning action uniformly,
users would hesitate to take such suggestions.
Consequently,
they still have to rely on DBAs' experience to make informed decisions.

Because of both the online and offline cost-optimization challenges,
users today typically struggle to control their cloud bills when using a data warehouse service.
We, therefore, envision the next-generation cloud-native system to be cost-intelligent.
The concept of \textbf{cost intelligence} is defined as the system’s ability to self-adapt to stay
\emph{Pareto-optimal} in the performance-cost trade-off under different workloads and user constraints.
As shown in \cref{fig:cost-intel},
a cost-intelligent data warehouse would (re)configure itself automatically
(e.g., through automatic cluster resizing and offline tuning) to move toward Pareto efficiency
so that users can easily make performance and cost trade-offs based on their application needs
by sliding along the Pareto frontier without worrying about wasting resources.

\begin{figure*}[t!]
\centering
\includegraphics[width=0.9\linewidth]{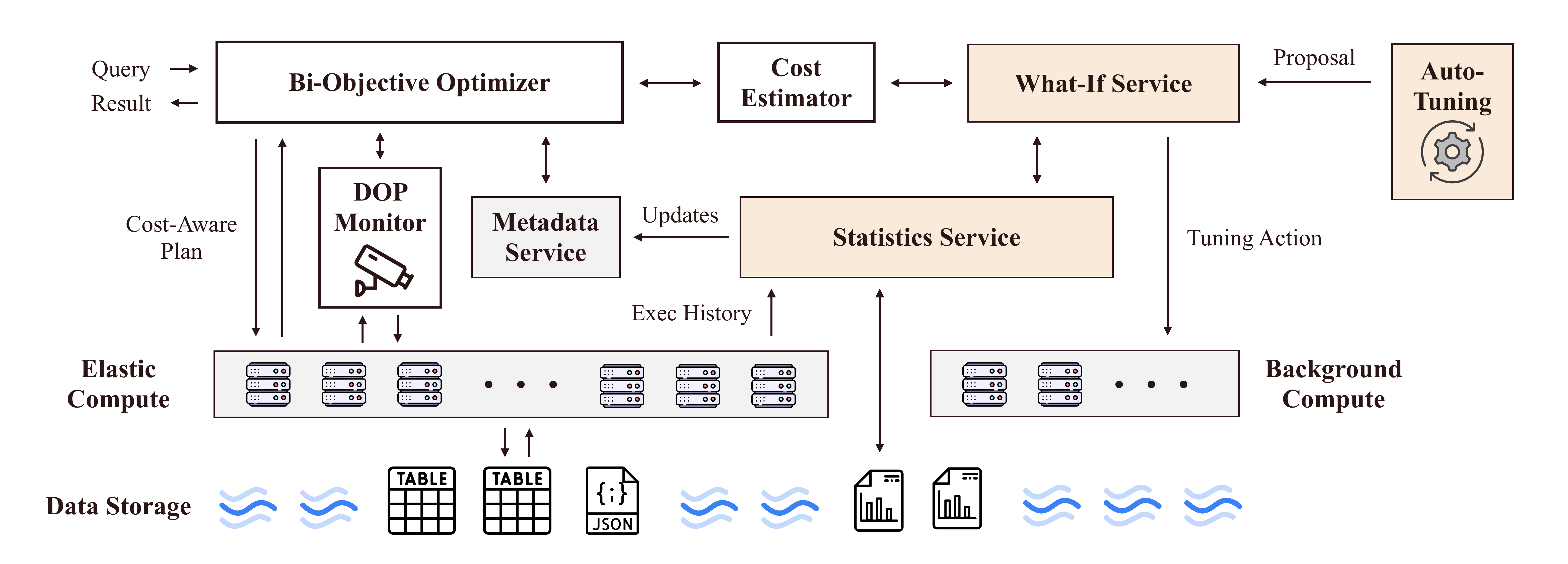}
\caption{System Architecture of a Cost-Intelligent Cloud Data Warehouse}
\label{fig:arch}
\end{figure*}

To run a workload in a cost-intelligent data warehouse,
users only need to specify their constraints/preferences on performance and cloud budget
(instead of a list of ``T-shirt'' sizes),
and the database would figure out how to deliver the query results on time with minimal dollar costs.
Additionally, each database tuning proposal is associated with a report that uses the dollar
benefit/cost as the bridge to evaluate the trade-offs
so that users can decide whether to apply this tuning action
without the need for expertise from professional DBAs.

We next sketch our solutions to the problems of
automatic resource deployment (\cref{sec:resource})
and cost-oriented auto-tuning (\cref{sec:tuning}).
We will present the architecture of a cost-intelligent cloud data warehouse
with an emphasis on the components and algorithms that are missing in today's systems.

%% file: resource.tex
\section{Automatic Resource Deployment}
\label{sec:resource}

In this section, we investigate the following problem:
given an analytical query, how many compute nodes should the cloud database allocate
to achieve minimal cost while satisfying the performance SLA
(or to achieve optimal performance within the cost budget).
We assume that the nodes are symmetric.
Selecting the optimal hardware configuration for a query is beyond
the scope of this paper, and we refer the readers to~\cite{leis2021towards}.

We adopt a basic system architecture similar to Snowflake
with disaggregated computing and storage~\cite{dageville2016snowflake}.
As shown in \cref{fig:arch}, the bottom storage layer,
hosted by cloud objects storage services such as AWS S3 and Azure Blob Storage,
keeps the user data in hybrid-columnar formats such as Parquet and ORC.
On top of that, the elastic compute layer allows users to acquire virtual machines on demand to execute database queries.
These nodes only provide computing power, and they do not hold any persistent states.
Such a disaggregated architecture enables the computation and storage resources to
scale independently.
At the top, the multi-tenant query optimization service parses SQL queries
and generates distributed execution plans for the compute nodes to work on.
There is also a metadata service that provides low-latency access to the
system's catalog and table statistics necessary for query planning.
In the rest of the paper, we assume private computations,
i.e., the (virtual) compute nodes are not shared among users.
We also assume that the database service provider maintains 
a warm server pool
to facilitate rapid cluster creation, resizing, and reclamation\footnote{Estimating
the warm-pool size at the service level is beyond
the scope of this paper.}.

As discussed in \cref{sec:motivation},
the diversity of database operators leads to a different cost-optimal
degree of parallelism for each pipeline within a complex query.
For example, scaling out a large table scan incurs less performance overhead
compared to scaling out a distributed hash join.
There are two types of methods to approach optimal DOP assignments.
The first is to determine the DOP of each pipeline at query optimization
(i.e., static planning).
However, no matter how sophisticated the models and algorithms are,
these static assignments could be far from optimal if the
cardinality estimation is way off.
On the other hand, the system can adopt a purely dynamic approach
where each pipeline starts with a default number of nodes (e.g., one)
and then gradually adjusts the DOP based on the real-time feedback on
the performance and resource utilization.
This approach, however, often leads to noticeable system overhead
caused by excessive cluster resizing operations.

We, therefore, propose a hybrid solution (shown in \cref{fig:arch}).
The initial DOP for each pipeline (i.e., execution stage)
is determined at query optimization time through static planning
that obeys user constraints on performance and budget.
This requires a relatively accurate time and cost estimation
at a fine granularity (\cref{sec:cost-estimation})
and an optimizer that considers both query latency and monetary cost
as it enumerates the plan space (\cref{sec:optimizer}).
Additionally, the system monitors the execution progress along with
the resource utilization and adjusts the DOP assignments at run time (\cref{sec:dop-monitor})
to ensure that the performance SLA is met with minimal cost
(or to obtain maximized performance under the budget).

\subsection{Cost Estimation}
\label{sec:cost-estimation}

The ability to predict the cost of a plan fragment accurately is indispensable
to a cost-intelligent system.
The cost estimator is at the center of our proposed architecture,
where it functions as a referee that ranks different execution proposals
(including background tuning tasks)
to guide the system to overall cost efficiency.
Here, cost refers to both time and monetary costs.
The input to the cost estimator includes both logical information such as
the plan shape and the input/output cardinality for each operator
as well as physical parameters such as
DOP assignments and other hardware characteristics (e.g., memory size).
The cost estimator then outputs the predicted time and monetary cost of
executing this plan fragment with the specified hardware configuration.
Because we assume private computation in the system,
the monetary cost of a workload is proportional to the total machine time
instead of the CPU time.
For example, if a pipeline execution is blocked on a node waiting for
the input data, the user is still charged for the under-utilized resources.

There are several desired properties of the cost estimator.
First, the \emph{accuracy} of the predictions determines the cost optimality
of the selected distributed execution plan.
A misleading cost estimation may cause serious under/over-provisioning
of the compute nodes and sub-optimal query plans (discussed in \cref{sec:optimizer}).
Second, the cost estimator must be \emph{lightweight}.
As the focal point of the architecture,
the cost estimator is frequently invoked by the foreground query optimizer
and the background tuning modules,
and its complexity affects the overall system efficiency.
Third, the models used in the cost estimator should be \emph{explainable}.
Because estimation errors are inevitable (due to cardinality misestimation),
the system would occasionally make bad decisions that lead to
slow and expensive execution of customer queries.
The explainability of the cost estimator allows database engineers
to reason about the root causes and propose fixes to make the module
more robust along the way.
When designing the algorithms in the cost estimator,
we are willing to trade some prediction accuracy for better
efficiency and explainability of the module
because most sub-optimal DOP assignments caused by the estimation errors
can be recovered by the ``DOP monitor'' at run time with a moderate
system overhead.

There is a rich literature on query performance prediction.
Although the algorithms proposed in prior work are inspiring,
they are insufficient for solving our cost estimation problem.
First, most prior models only predict job/workload-level performance
through learning the high-level job execution patterns
\cite{venkataraman2016ernest,pimpley2022towards,ferguson2012jockey}.
They, however, are ignorant of the pipeline formation inside a complex query
and are thus unable to provide more fine-grained cost estimations.
Second, these models target big data systems with a simpler map-reduce execution model.
Compared to executing analytical queries in a relational database,
map-reduce jobs do not involve direct data transfer between execution nodes,
and they do not have pipelines running in parallel within a job.

Many previous solutions rely heavily on machine learning~\cite{ganapathi2009,tang2021forecasting,akdere2012learning,pimpley2022towards,marcus2019,fan2020comparative}.
There are several issues with this approach.
First, they typically assume a recurrent workload (common in big data systems)
and train their models offline using features extracted from the execution history
\cite{pimpley2022towards,ferguson2012jockey,siddiqui2020cost}.
These models, however, may not be generalized enough to provide accurate performance
predictions for ad-hoc queries issued by data scientists in a data warehouse.
To obtain a robust performance prediction for an arbitrary query,
many performance prediction algorithms choose to profile the query on small data samples
to collect training data for their ML models~\cite{venkataraman2016ernest,rajan2016perforator,karampaglis2014bi}.
Although the overhead of the training process can be bounded~\cite{venkataraman2016ernest},
it is unlikely that such a sampling-based estimation is lightweight enough
to be invoked frequently during query planning and execution to adjust the DOP assignments.
Furthermore, ML-based performance predictors often feed high-level query features extracted
from the SQL text (e.g., word frequencies) and physical plan (e.g., number of operators) 
into models such as XGBoost and SVM~\cite{tang2021forecasting,ganapathi2009, akdere2012learning,fan2020comparative}.
These models treat the query internals as a black box
and thus compromise the prediction explainability,
a property desired in our architecture.

Designing a cost estimator that is accurate, lightweight, and explainable
is a challenging research problem.
In the rest of this section, we sketch a possible solution that we are actively investigating.
Our cost estimator contains a set of per-operator models
and a query-level simulator.
For each physical operator, we design a scalability model that outputs
its processing throughput given the data size and the degree of parallelism.
The model also refers to the relevant hardware parameters that are calibrated
before the service starts.
We found that simple mathematical formulas are good enough to model
the scalability of most physical operators (e.g., scan, filter).
To improve the prediction accuracy for more complex operators
(typically involve data exchange between nodes),
we pre-train regression models for them with synthetic workloads that
cover the parameter space.
As discussed before, we try to avoid using complex ML models (e.g., deep neural networks)
that trade explainability for further accuracy.

Based on the per-operator scalability models,
we can compute the throughput of an operator pipeline given a DOP assignment
and thus estimate its execution time and total machine time ($\propto$ cost).
The query simulator then models the data flow in each pipeline of a query plan.
In a multi-pipeline query, pipelines could be executed in parallel, and a downstream
pipeline could be blocked if the data from one of its parents is not ready.
The query optimizer, therefore, would invoke the simulator multiple times
to find a cost-optimal pipeline-level DOP assignments
(e.g., the accumulated ``blocked'' time of the pipelines is minimized).

\subsection{Bi-Objective Query Optimization}
\label{sec:optimizer}

Query optimization in a cost-intelligent cloud data warehouse must be bi-objective.
The optimizer receives user constraints (or preferences) on query latency and cloud budget
and produces distributed query plans that are most efficient while satisfying these requirements.
The bi-objective optimizer invokes the cost estimator discussed in the previous subsection
to be cost-aware when searching the plan space.
The key challenge in designing a bi-objective optimizer is to keep its computational
complexity low.

Previous studies proposed theoretical frameworks for solving the multi-objective optimization
problem in databases~\cite{trummer2014approximation}.
These solutions target producing a set of physical plans that form the Pareto frontier
of the trade-offs of the multiple objectives.
However, generating a series of optimal plans with different trade-offs inevitably
adds significant computational complexity to the search algorithm.
We argue that it may not be necessary for a cost-intelligent database to present
the full spectrum of plans with different estimated times and costs for users to choose from.
We observe that it is more friendly to users to directly specify their latency or budget
constraints for a query.
Therefore, we can ``downgrade'' the bi-objective optimization problem into
a constrained single-objective optimization problem
(i.e., find a plan with a minimal monetary cost that satisfies a latency requirement,
or find a plan with minimal query latency within a cloud budget)
to achieve a search complexity similar to a traditional cost-based optimizer.

A second source of complexity unique to our optimizer is the fine-grained DOP planning.
Ideally, DOP planning should be integrated into the unified cost-based search~\cite{graefe1995cascades}
to obtain an optimal distributed plan~\cite{viswanathan2018}.
However, enumerating the DOP for each pipeline while exploring the physical plan shape
makes the search space explode.
Instead, we separate the DOP planning from the DAG planning
(i.e., the traditional single-machine query optimization that produces an execution DAG)
into a subsequent optimization stage.
Specifically, searching for an optimal DOP assignment only applies to the ``chosen'' plan
produced by the DAG-planning stage.
Although the separation of the stages misses the opportunities to reach a globally optimized plan
obtainable from a unified search,
it keeps the search complexity comparable to existing optimizers.

Because a pipeline cannot start until all of its dependent pipelines are complete,
a heuristic that we use to speed up DOP planning by pruning the search space
is to make sure that these (concurrent) dependent pipelines finish roughly at the same time
to minimize resource waste due to pipeline waiting.
Specifically, if the two dependent pipelines started at the time have input cardinalities $C_1$ and $C_2$,
and the throughput functions given by the cost estimator are $T_1(\cdot)$ and $T_2(\cdot)$,
we ensure that the DOP assignments of the two pipelines satisfy
$\frac{C_1}{T_1(DOP_1)} \approx \frac{C_2}{T_2(DOP_2)}$.

A bi-objective query optimizer in a cloud system should leverage resource elasticity
to make judicious trade-offs between query latency and monetary cost.
Optimizing bushy joins is one of the most interesting problems in this area.
In a distributed environment with elastic resources,
a ``bushier'' plan enables more concurrency in pipeline executions
and is more likely to have a lower query latency.
However, a bushier plan may not be optimal in terms of join cardinalities,
and it may, therefore, cost more computations (and total machine time).
Bushy joins are usually ignored in traditional optimizers for single-machine
(i.e., the DAG planning stage in our optimizer) to reduce the search space.
We propose to explore bushy plans in the DOP planning stage.
After receiving a left-deep plan from DAG planning,
we would reorganize the join shape to make a series of plan variants
that are increasingly bushier.
The relations are chosen carefully in the above plan rewrite
so that the join cardinalities are bounded (e.g., non-expanding joins)~\cite{chen2016memsql}.
We then apply DOP planning to each of the plan variants and choose the one
that makes the best time-cost trade-offs under user constraints.

\subsection{Dynamic Cluster Resizing}
\label{sec:dop-monitor}

A static DOP assignment produced in query optimization could suffer from
errors in cardinality estimations.
We, therefore, introduce a DOP monitor that dynamically adjusts the cluster size
at run time to meet user requirements.
Prior auto-scaling strategies typically fall into two categories.
The first is to assess the execution progress after each fixed time interval
and scale the cluster if necessary to meet a performance SLA
~\cite{ferguson2012jockey,thamsen2017ellis}.
This approach works well for massively parallelizable jobs (e.g., map-reduce).
For complex analytical queries, however, scaling out the entire cluster
may not be cost-efficient.
For example, if the execution is bottlenecked by a particular pipeline,
scaling out the concurrent (or downstream) pipelines proportionally could
hurt their resource utilization.

Therefore, we apply auto-scaling at the pipeline granularity.
The DOP monitor collects the true cardinalities, the pipeline flow rates,
and the resource utilization at run time.
If the measures of a pipeline deviate from the statically-planned values
within a threshold, we correct the deviation by adjusting the DOP of
this pipeline only (according to the scalability models in the cost estimator).
If the deviation is substantial,
we will reinvoke the DOP planner with the collected run-time statistics
to generate a new set of DOP assignments for all the pipelines
to ensure that user constraints are satisfied efficiently.

The second category of auto-scaling strategies is to determine the resources for
the next execution stage after each data shuffle~\cite{siddiqui2020cost,yint2018,bigquery}.
For example, Google's BigQuery~\cite{bigquery} would shrink the cluster size for
the next stage if its shuffle service detects severe overestimation of the
output cardinality of the previous stage~\cite{edara2021big}.
This approach relies on materializing the intermediate results at the pipeline breakers
on persistent storage or in a data shuffle service.
Such ``clean cuts'' between execution stages impose performance overhead,
and we believe that they are nonessential to achieving fine-grained auto-scaling.
Our DOP monitor can not only re-plan the DOPs for future stages but also
adjust the cluster size of the current stage with minimal resizing overhead.
This is enabled by the morsel-driven scheduling~\cite{leis2014morsel} in our execution engine,
where the smaller tasks make real-time cluster resizing more efficient.
We also adopt a push-based execution model~\cite{duckdbpushbased} so that we have centralized control
over the data flow to allow DOP changes promptly.

%% file: tuning.tex
\section{Cost-Oriented Auto-Tuning}
\label{sec:tuning}

As discussed in \cref{sec:motivation},
a key step toward automating the database tuning process in the cloud
is to use the \emph{monetary cost} as a common metric to evaluate
different aspects of a tuning action systematically.
Physical database tuning is a difficult problem in traditional
DBMSs with fixed resources.
It relies heavily on DBAs' experience
because the resource contentions are hard to quantify.
For example, creating a materialized view (MV) for an intermediate
join result would speed up a group of queries and thus improve the
read throughput of the system.
However, maintaining the freshness of the MV would slow down
writes to the database.
And because of fixed resources, spending extra computation on MV updates
also hurts the system's read-throughput.
This implicit resource contention between the read and write operations
complicates the tuning process.

We argue that the auto-tuning problem is more solvable in a cloud environment.
The key idea is to leverage the elastic resources to guarantee the same or
better performance after applying a tuning action
and then evaluate whether this action reduces the operational cost of the system
in the long run.
Consider the same MV-creation example above,
we would allocate separate compute resources for MV maintenance to avoid
resource contention so that it does not hurt the performance of normal
read and write operations.
Then, we estimate that the computation saved by substituting the MV into queries
is worth $x$ dollars per time unit, and the extra cost of storing and updating
the MV is $y$ dollars per time unit.
If $x - y > 0$, this tuning action is likely to be beneficial.
Using the dollar as a common metric simplifies the auto-tuning logic
and makes it possible to present the trade-offs of a tuning action
to average customers clearly.

To estimate the above dollar benefits/costs for a tuning action,
the system must be able to predict future workloads besides accurate cost estimation
described in \cref{sec:cost-estimation}.
Recent work has focused on using various machine-learning algorithms for the task~\cite{ma2018query}.
Although these algorithms matter, we believe that a comprehensive and efficient
\emph{Statistics Service} (as shown in \cref{fig:arch}) is the foundation of accurate
workload predictions.
For each database instance, the Statistics Service collects the query execution logs
from all the tenants to form the ``ground truth'' for understanding workload behaviors.
The service computes in the background with these collected traces to generate
and maintain queryable workload summaries,
including file/attribute-access counts and weighted join graphs\footnote{A graph
where the vertices are table attributes and
the weights on the edges indicate how often the attributes are joined.}
for training workload-prediction models
and run-time resource usage for modeling the performance and monetary cost.

We identify several challenges in building an efficient Statistics Service.
First, the database must implement its own lightweight profiling tool that
can attribute the run-time resource measures to logical database tasks easily.
Most off-the-shelf profiling tools, such as Linux Perf, incur prohibitively high
system overhead when accompanying normal query execution,
and they are only good at capturing snapshots of entire processes.
Second, although we assume private computing at the user level,
it is multi-tenant underneath for the service provider.
The ability to attribute the shared hardware usage to each of the
concurrent workloads are critical to the Statistics Service.
Finally, the Statistics Service itself must be cost-efficient as well.
This requires new algorithms to balance the generation cost and
the comprehensiveness of the statistics (e.g., by varying sampling rates).
The service could identify the hot and cold statistics
and design different data structures on tiered storage
to trade-off between querying performance and storage cost.

To complete the auto-tuning cycle, our system includes a What-if Service
that evaluates tuning proposals from existing auto-tuning tools~\cite{chaudhuri1998autoadmin,ma2018query}.
For each tuning proposal, the What-if Service generates a relevant workload
prediction based on the Statistics Service.
Then it invokes the cost estimator to determine whether the tuning action
is ``profitable'' using the logic described at the beginning of this section.
Once the What-if Service accepts a tuning proposal
(the process could involve user approval),
the job is sent to the background compute for execution.

%% file: concl.tex
\section{Conclusion}
\label{sec:concl}

We introduced the concept of cost intelligence,
a much-desired property for next-generation cloud data warehouses.
The architecture proposed in the paper allows
both automatic resource deployment and cost-oriented auto-tuning.
We are actively building the system at Tsinghua University,
and we hope that the paper will inspire talented researchers
in the community to tackle the challenges presented in the paper together.